# Narralive – creating and experiencing mobile digital storytelling in cultural heritage

Ektor Vrettakis, Vassilis Kourtis, Akrivi Katifori, Manos Karvounis, Christos Lougiakis, Yannis Ioannidis

Department of Informatics and Telecommunications, University of Athens & Athena Research Center

## Abstract

Storytelling has the potential to revolutionize the way we engage with cultural heritage and has been widely recognized as an important direction for attracting and satisfying the audience of museums and other cultural heritage sites. This approach has been investigated in various research projects, but its adoption outside research remains limited due to the challenges inherent in its creation. In this work, we present the web-based Narralive Storyboard Editor and the Narralive Mobile Player app, developed with the objective to assist the creative process and promote research on different aspects of the application of mobile digital storytelling in cultural heritage settings. The tools have been applied and evaluated in a variety of contexts and sites, and the main findings of this process are presented and discussed, concluding in general findings about the authoring of digital storytelling experiences in cultural heritage.

## Introduction

Storytelling has the potential to revolutionize the way we engage with cultural heritage and has been widely recognized as an important direction for attracting and satisfying the (ever more demanding) audience of museums and other cultural heritage sites (Pujol et al., 2013; Bedford, 2001; Twiss-Garrity et al., 2008). Nevertheless, it comes with a unique set of challenges, which are not present either in traditional storytelling or more information-driven guided tours.

Paradoxically, the first of these challenges is to define it clearly, as it is common practice for a variety of multimedia experiences in cultural heritage settings, including guides offering purely informational content, to be self-characterized as storytelling. Taking from the work of (Katifori et al., 2018), we define storytelling in a cultural heritage site as an experience built around a continuous, coherent narrative that leverages the interpretation of the available cultural heritage artefacts to develop the essential elements of storytelling: setting, characters, plot, conflict, theme, and a satisfying narrative arc (e.g., setup, tension, climax, and resolution). This experience is delivered at and meaningfully staged on a cultural heritage site and may or may not directly correspond with real events.

This approach has been investigated in different research projects like CHESS (Katifori et al, 2014) and EMOTIVE (Katifori et al, 2019) but its adoption outside research remains limited. We believe this is not due to a fundamental disparity between storytelling and cultural heritage, but due to the fact that mobile digital storytelling for heritage encompasses all the challenges of good storytelling, extended by the spatial and interpretative constraints of the cultural heritage site it must be applied on.

Furthermore, as (Katifori et al., 2018) note, there is a distinct lack of experts with the experience to design for this particular medium. Based on our previous work (Roussou et al, 2015), the design of digital storytelling experiences for archaeological museums (and probably for cultural settings in general) is a multi-layered creative process requiring interdisciplinary authoring groups. These usually include members with complementary skills, which fall under the following general categories:

- Archaeologist: informational content, interpretation
- Museologist - Curator: visitor needs analysis and development of the story in relation to the gallery and the exhibits
- Museum educators: visitor needs analysis, development of the story taking into account the specific needs of the younger audience
- Storyteller: story concept and draft scenario
- Creative designer: asset creation, story production

The institutions mostly in need of attracting larger audiences through innovative applications, i.e., low-profile museums and cultural heritage sites, also lack the necessary funds and experience to put together such a team and coordinate the overall effort.

To address this gap in research, in this work we present the web-based Narralive Storyboard Editor (NSE) and the Narralive Mobile Player (NMP) app, developed with the objective to assist the creative process and promote research on different aspects of the application of mobile digital storytelling in cultural heritage settings. The Narralive tools offer the following:

- An authoring methodology that uses storytelling structural elements (chapters, scenes, pages) to reinforce the importance of narrative and indirectly guide the author to a story-centric approach.
- A set of templates for presenting content on the visitors' mobile devices without having to code or decide on complex UI design / post-editing issues (e.g., how should an image be presented depending on its aspect ratio and resolution based on the screen size, resolution, and orientation of the visitor's mobile device). These templates range from simpler ones such as presenting images, text, and audio, to more complex ones, especially relevant for storytelling, like fully voiced-over dialogue among story characters.
- An intuitive, simple and cost-effective, web-based solution that can be used by authors with different levels of expertise to easily draft and implement their story concepts and evaluate them on-site.
- A light-weight, stable Player app, compatible with most mobile devices, to consume digital storytelling content.

Our main objective is to create an easy-to-use tool that can be used widely to test digital storytelling experiences in real life contexts and thus support research and experimentation in visitor experience design for heritage.

In the sections immediately following the introduction, we compare our work with other relevant authoring and experiencing solutions and then briefly describe the NSE and NMP tools. Next, we focus on the use and evaluation of these tools in a variety of contexts and summarize our main observations, followed by a more general discussion concerning the application of digital storytelling in cultural heritage. We conclude the paper in the last section.

# Digital storytelling authoring in cultural heritage

The Narralive Storyboard Editor was designed and developed after identifying a gap in technological solutions for museum and cultural heritage sites that wish for a cost-efficient way to create and offer digital storytelling experiences. Arguably, there is a variety of existing tools for the authoring of guided tours and other similar experiences for cultural heritage, or narrative experiences in other contexts.

More specifically, a large category of tools follows a "point of interest" approach, generally appropriate for guided tours in cultural heritage. The GuidiGO mobile app and the accompanying web-based authoring tool GuidiGO Studio (Guidigo, 2019) support the creation of text and multimedia content with the possibility for some more interactive elements like challenges or quizzes. These templates serve the purpose of creating a gamified, location/exhibit-based experience and cannot be easily applied to complex story plots. Another application offering similar capabilities is the izi.TRAVEL CMS (izi.travel, 2019), which supports content in the form of text, images, audio, video and quiz, appropriately offered at specific locations on a map/floorplan, thus connecting them with points of interest and exhibits. The author can set these locations either as a linear experience with a specific order of visit or as an experience where the user freely visits each one of them. This tool makes no provision for branching narratives and is agnostic of any overarching plot.

Another category of tools comes from the interactive storytelling domain and focuses on complex branching story structures, usually visualizing them as a graph of script parts as nodes and their possible connections as edges. Taking into account the Interactive Digital Narrative authoring tools classification work in (Shibolet et al, 2018) and accompanying IDN Authoring Tools Resource appendix, our tool falls within the area of interactive fiction tools and more specifically within the "Hypertext" category. A popular representative of this category is Twine (Twine, 2019), and the CHESS authoring tool (Vayanou et al, 2014) is an example applied to cultural heritage. Usually these tools export the end result as a form of interactive hypertext, possibly accompanied by audio or images.

In conclusion, the examined existing authoring tools offered to cultural heritage sites are more suitable for creative designers with technical expertise, than to the rest of the identified author categories. Furthermore, they either focus on a 'points-of-interest' approach or focus on the support of complex branching narratives without offering particular design guidance for incorporating storytelling into experiences in a meaningful and intuitive way.

Our Editor was designed with a story-centric approach in mind, where the structure of the experiences is driven by the plot and not the locations of the points of interest and exhibits. A storyboard-like look-and-feel and a hierarchical organization of the contents in story terms (chapter, scene, page) were combined with additional structural elements to support branching in the narrative, boosting plot interactivity and engagement. Furthermore, we provide content templates that go beyond images and audio, resulting in a final experience that is distinct from and beyond hypertext. The next section provides an overview of the main characteristics of the NSE.

# The Narralive Storyboard Editor

The Narralive Storyboard Editor (NSE) is a web-based authoring tool for digital interactive narratives. Its purpose is to aid users that possess storytelling skills but lack strong technical or creative expertise to lay down experiences in an easy and effective manner. Interactive experiences are not necessarily linear, but may contain decision points or optional parts. Hence, they are mandatorily modelled as graph structures. The problem posed for most authors is that graph structures as a representation are counter-intuitive and hard to understand. NSE abstracts this structural complexity thus allowing the author not to think in abstract terms such as graph nodes and edges. Instead, we introduce a set of basic storytelling structural primitives such as chapters, scenes and pages that provide an easy and intuitive way to create, organize, visualize and alter the experience content and structure. These primitives are backed by predefined sets of templates for presenting content and alternative paths of the main story plot. In order to mask the complexity of structuring an interactive experience while at the same time aiding the authoring process, NSE adopts the storyboard paradigm. A storyboard is a graphic organizer in the form of illustrations displayed in sequence for the purpose of pre-visualizing an interactive media sequence. The storyboard-like look and feel helps to organize structural and content primitives in a sequential, visually appealing graphical manner.

An experience is composed of its structure and content. The content is everything that the end user experiences through the Mobile Player while structure defines the story plot. To organize the structure and content of an experience, NSE enhances the storyboard paradigm with a stacked architecture of layers which consists of the following building blocks (Figure 1):

- **Stories** represent complete experiences and act as containers for the rest of the layers.
- **Chapters** organize large, coherent parts of the story plot.
- **Scenes** further break down chapters into smaller, self-contained plot pieces that are characterized by coherency in time, space or characters.
- **Pages** are self-contained components that house all the content that the user experiences. The content presentation is backed by a number of predefined screen templates.

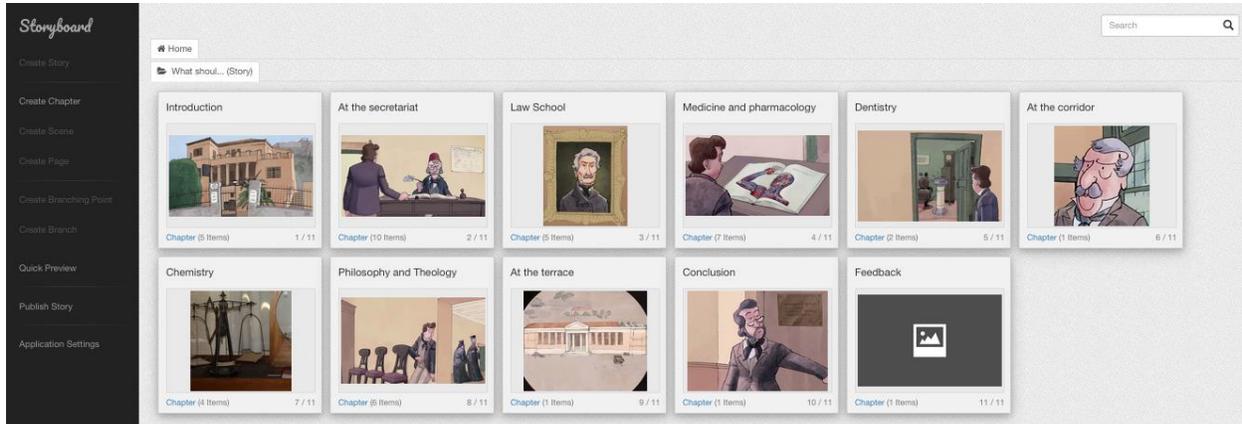

Figure 1. The sequence of Chapters in a Story, as presented in the NSE

The author has to use **Menu** elements to specify alternative paths of the plot. Menus exist within Chapters or Scenes and can be nested indefinitely. Menus can be of type **More** or **Choice** and are also backed by templates. A menu of type Choice represents a unique turn of events in the story plot, whereas a Menu of type More aims to provide the user with the option to view additional content. In Figure 2, nested Choice and More menus are presented. When the user selects a menu, the menu expands in order to show its contents.

Templates allow the author to structure the desired content without having to code or decide on complex UI design as the platform creates visually appealing UI. For this purpose, NSE offers a wide variety of predefined templates that range from simple ones, used to present text, images and audio to more complex ones, especially relevant to storytelling and/or cultural heritage, like fully voiced-over dialogue among characters. The NSE content template types include the following:

- **Simple**: A template for creating a screen consisting of rich text, images and audio clips.
- **Voiced-Over Dialogue**: A template for creating a screen that embodies voiced-over dialogues between a set of characters.
- **Quiz**: A template for posing Right or Wrong Questions to the user.
- **Video**: A template for creating screens that contain video clips.
- **Interactive Image**: A template that allows the author to define hotspots on an image as well as interactions with them (i.e. rendering of some text or the playback of a sound).
- **Interactive Book**: A template that consists of multiple Interactive Images and aims to emulate book skimming (Figure 3).
- **NFC Reader**: A template for prompting the user to scan a NFC tag in order to trigger some interaction.
- **Question**: A template presenting a questionnaire for the user to provide textual feedback.

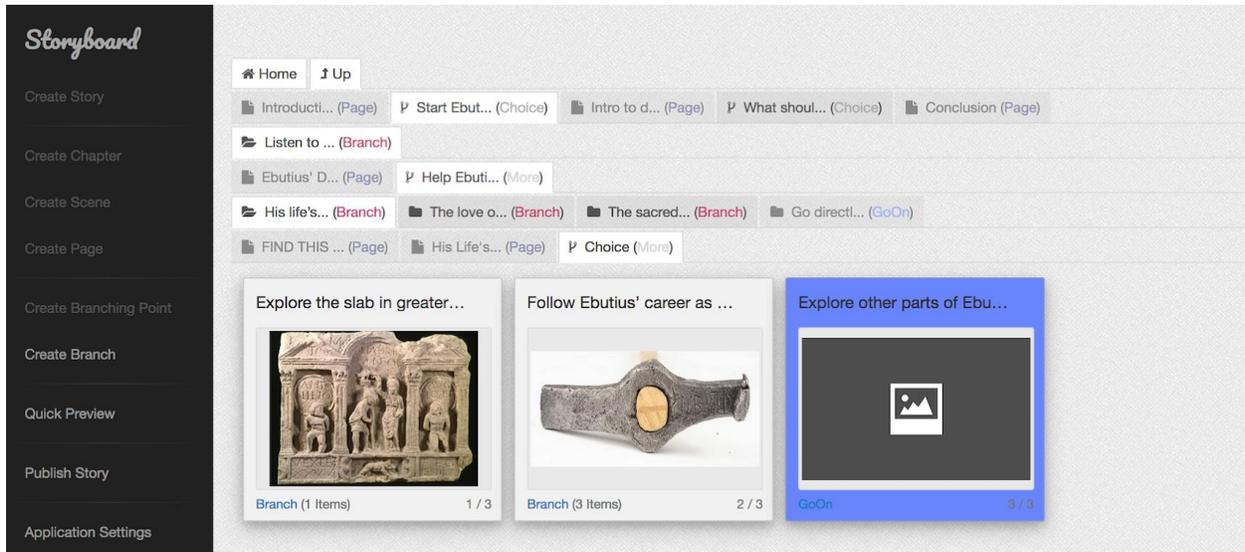

Figure 2. Nested menus

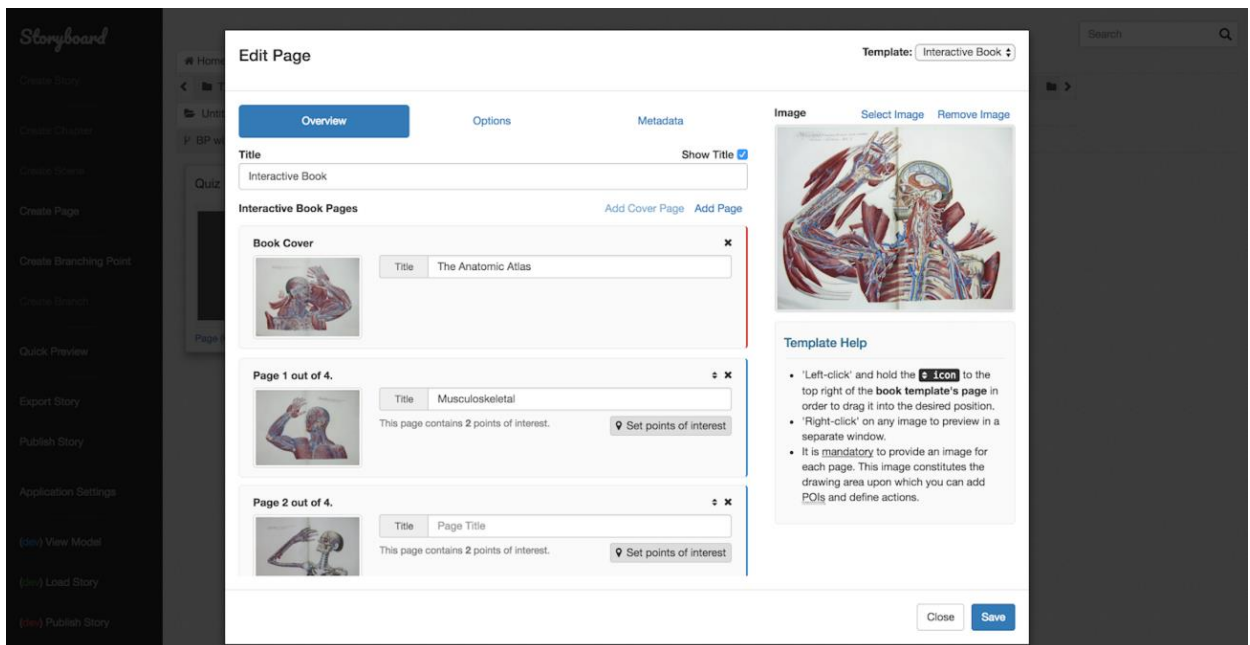

Figure 3. The NSE interactive book template. In this example we see the anatomic atlas at the University of Athens History Museum, with the book cover and 2 out of 4 pages of the book listed. As we can see the author can specify the title of the pages as well as points of interest on the page.

Moreover, NSE offers the following Menu templates:

- **Tiles / List**: A listing of the alternative paths the user may choose to follow, styled as tiles or list elements.
- **Interactive Image**: A Menu template that represents alternative story paths as points of interest (POIs) on an image. By selecting a POI, the user triggers the selection of the corresponding path.
- **Map**: The author matches alternative story paths with trigger areas on a map. When the user's mobile device physically enters such an area, the corresponding path is triggered.
- **QR-Code**: The author matches alternative story paths to QR-Codes. When the user scans one of the NSE-generated QR-Code with the Narralive mobile app, the corresponding path is triggered.

The NSE offers specific key functionalities for editing and publishing experiences. Its preview functionality allows the author to view and examine the story at any point in time during its creation, without exporting it to view it on the Mobile Player. It enables the author to fine-tune the story and make adjustments by emulating the story structure traversal and content presentation. The preview is a live component, meaning that changes are immediately visible to the author. NSE also provides basic editing capabilities such as copying and moving story elements. Making the experience available to the Mobile app is straight forward through the Publish functionality.

The NSE is written in JavaScript on top of the angular framework and backed by a document store. Any communication between the app and the backing store is made through a RESTful interface.

The experiences created with the NSE are delivered to the end users through the Narralive Mobile App, which is described in the following section.

# The Narralive Mobile Player

The Narralive Mobile Player (NMP) is a mobile application that supports the delivery of experiences, created with the NSE, to the visitors. It is designed with simplicity and intuitiveness in mind and allows visitors to explore, download and "run" the available experiences. The app is easily customizable, at development time, to support different needs of cultural heritage sites, such as the visual identity (e.g. app's icon) and the general look and feel. This feature becomes prominent on the following screenshots (Figures 4 to 7) where one can see the various "flavours" that have been used in various instantiations of the NMP.

One of the basic features of the app is exploring and downloading experiences, which constitutes the "dashboard" functionality of the app. The NMP communicates with the backing store and populates a list of available experiences. Depending on the app's "flavour" the visitors will see a list of the available experiences for a particular cultural heritage site. Selecting an experience leads to the experience's screen where the visitors can read a description of the experience and download it (Figure 4). After downloading, the experience becomes available and the visitor is able to start it. If there are updates in the experience, it is possible to update it from the same page (Figure 5).

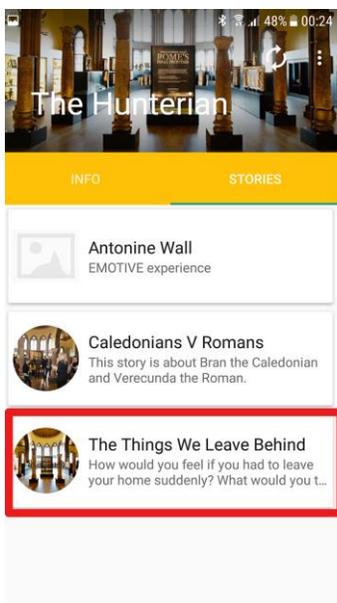
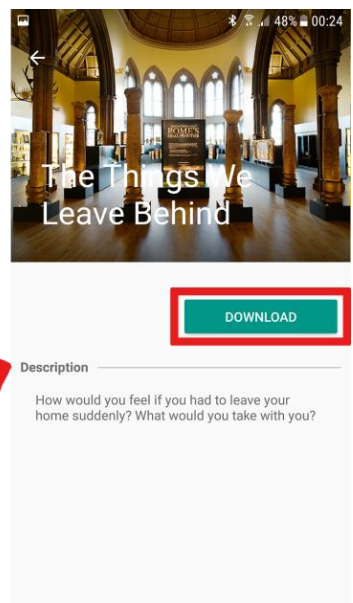
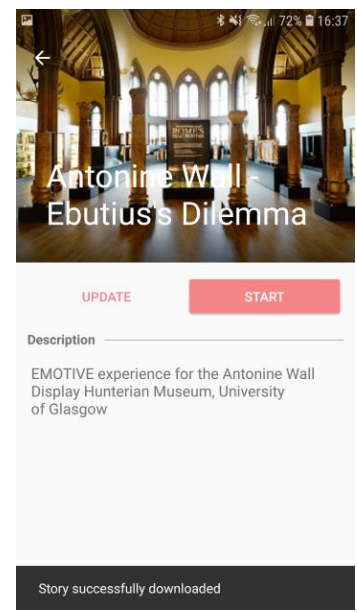

Figure 4. Experience selection and downloading            Figure 5. Downloaded experience

When an experience is started, the app enters the "presentation" mode. Through the visualization of the predefined Page templates, described in the previous section, the mobile application is able to present static or interactive activities to the visitors, based on the author's choices. The provided activities vary from simple pages with image and text, to quizzes, interactive images, voiced-over dialogues or questions directed to the users (Figure 6). Additionally, the experience may follow different paths, if the author chooses to introduce menus (branching points) into the narrative. In these cases, the visitor will see menus like the ones presented in Figure 7.

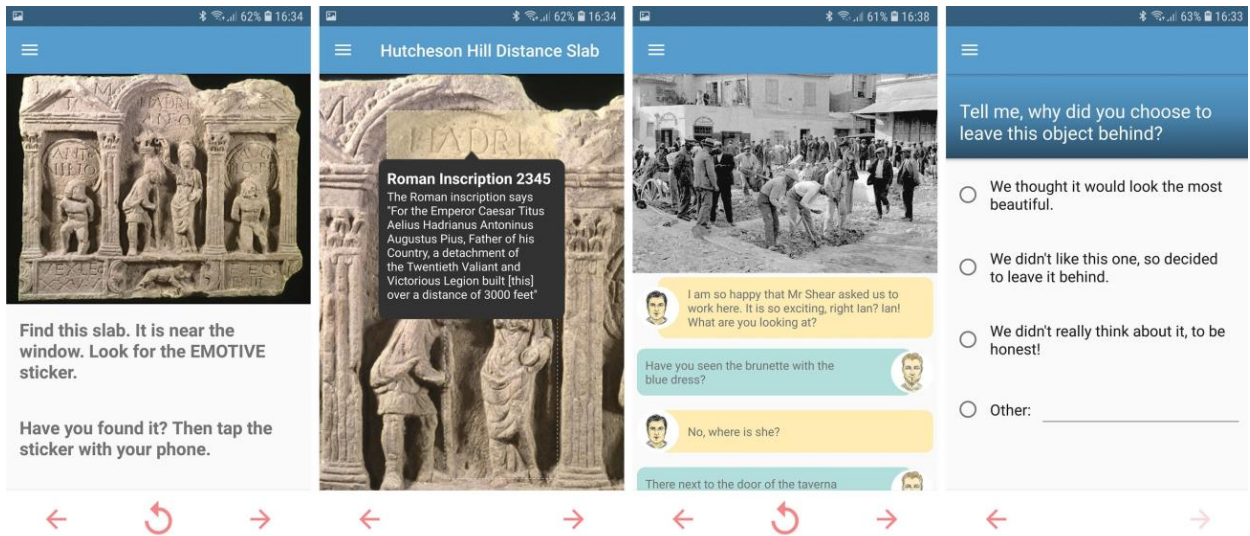

Figure 6. Some of the Page templates offered by the Narralive Mobile Player app for the visitors

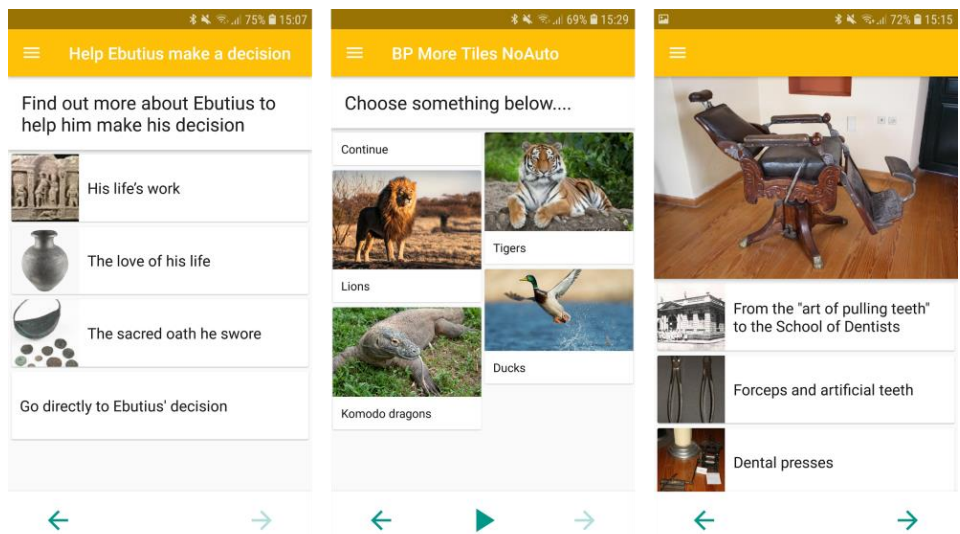

Figure 7. Examples of Menus where user have the ability to choose paths in the experience (List & Tiles templates)

The Narralive Mobile App is made for Android OS and works on devices featuring Android v4.1 (API 16) or later. It was developed with Android Studio, using Java and Android SDK, and utilizes Android Flavors to enable easy UI customization. An iOS version is currently under development and will become available late 2019.

# Narralive in practice

The Narralive Storyboard Editor tool has undergone several cycles of formative and summative evaluation since its design and development stages. Initial assessment in a controlled environment, monitoring its use by giving the users specific tasks to complete, was followed, in the alpha version, by longitudinal evaluation where users developed their own experiences and were asked to report issues and comments on the tools in regular intervals, through brief interviews.

Once the beta version of the tool was complete and tested, it has been made available to design experiences in real contexts. Our team provides continuous support in these cases to gather user feedback on issues with the tool and possible additions and improvements. As in the case of the longitudinal evaluation of the alpha, the users were regularly interviewed during the story development period with a final interview at the end.

The Narralive Storyboard Editor and Mobile Player have been put in practice for different purposes in a variety of contexts and use cases, within the course of the last three years. These experiences have been used for educational purposes, in the context of digital archaeology and museology, they support research in the field of museums and archaeological sites experiences, and they are in use by museums offering visits to the public. These different applications of our approach have helped us improve it in terms of user experience and to identify additional requirements for new functionality and activity templates.

In this section, we focus on this, last stage of our evaluation activities for the beta version of our tool in real life contexts. We provide an overview of these contexts, followed by the types of experiences produced and, lastly, we summarize the main outcomes.

# Contexts of use and evaluation

The NSE has been presented to its potential authors in a variety of Workshops and Events and it has been used by different cultural heritage institutions to author digital storytelling experiences for their sites. In all cases we recorded the author feedback on the NSE as well as visitor feedback on the NMP, including how visitors perceived the produced experiences. In this section we present briefly some of these evaluation activities, summarized in Table 1.

## DialPast Course Digital Storytelling Workshop

In the context of a DialPast Course that took place in Athens on 29 August to 3 September 2016 at the Norwegian Institute, course attendees participated in a half-day hands-on workshop where they were asked to use the NSE to create a storytelling experience for the Acropolis Museum Archaic Gallery. A total of 13 individuals from 10 different European universities participated in the workshop, PhD students in their majority, with an explicit research interest in digital archaeology.

For the authoring exercise, the users had been divided into four groups and had been asked beforehand to prepare the concept of their story using a simple worksheet. During the workshop, after a brief demonstration of the tool, they were asked to implement their story concept. An evaluator was present in each group, to offer support with the use of the tool and record comments, difficulties or any type of interesting issue. At the end of the Workshop, each group presented their story, and they were asked to provide their feedback in a focus group discussion and with individual questionnaires.

## Museum studies "Museums and New Technologies" postgraduate course

December to February 2016 & 2017 and February to June 2019 we supported the National and Kapodistrian University of Athens course "Museums and New Technologies" in the context of the postgraduate "Museum Studies" program by giving several lectures on digital mobile storytelling in cultural heritage and supporting the students to create their own digital storytelling apps in the context of their course project. The assigned museum for each project was selected from the list of the University of Athens museums. This process produced useful feedback both for the story authoring and experiencing.

45 students in total worked with Narralive, aged between 23 and 51, from diverse backgrounds with their bachelor mainly related to archaeology, however in some cases also with an engineering background. They formed an interesting mix of technological and theoretical expertise that led to a variety of approaches in story creation. The students formed groups of 2-4 and developed their digital storytelling project with the NSE after a two-hour hands-on tutorial. They were supported upon their request if they encountered issues during the creation of their stories. They then validated their complete prototype experiences in the museum they had selected.

### The EMOTIVE research project

The NSE has been one of the main authoring tools used in the course of the EMOTIVE project by the internal User Group of the EMOTIVE consortium and has resulted in the first prototype of the Hunterian Antonine Wall on-site experience, "The life of Ebutius", which was evaluated at the Scottish European Researchers' Night, Explorathon 2017 in the Hunterian Museum, on 29 September 2017 (Economou et al, 2019).

The formative evaluation of the tool is on-going, within the project, as in parallel with the Hunterian Museum experience, the tool has been used to create the Çatalhöyük Neolithic Site experience which was tested both off site and on-site at Çatalhöyük in the summer of 2017 and 2018 and also transformed to the Çatalhöyük Exploration of Egalitarianism Digital Classroom Kit[1]. This testing aimed to gauge both how visitors interacted with the NMP, as well as the appropriateness of the NSE to support the authoring of the experience (Mirashrafi, 2017: 41).

### "Real world" apps

The NSE has been used to create digital storytelling apps for the museum visitors in the case of two museums so far, the University of Athens History Museum and the Criminology Museum. In both cases the storytelling apps have been created through the collaboration of the museum curators, domain experts as well as creative designers and storytellers. In all cases an iterative design process has been followed, alternating design and development with evaluation activities. In the case of the History Museum, the app, named "What should I study", is currently available in Google Play Store for download and also to visitors on demand with mobile devices provided by the museum. In this case, evaluation data is collected through a short feedback form embedded in the app and through brief interviews with randomly selected visitors. An English version of the story will be available soon.

Table 1. Summary of NSE evaluation activities

| Activity | Period | Experts involved |
|---|---|---|
| DialPast Course Digital Storytelling Workshop | 2-day workshop, September 2016, Athens | 13 archaeology and museology experts |
| "Museums and New technologies" course, | November to February, | 30 museology |

---
[1] http://athena.emotiveproject.eu/schoolkit/

| "Museum studies" master program | 2016 and 2017 | postgraduate students |
|---|---|---|
| Experiences designed for research purposes (including for the EMOTIVE project), in collaboration with specific heritage sites | On-going since 2017 | 9 archaeologists, heritage experts, creative designers |
| Experiences designed in collaboration with the site experts to become available to the public | On-going since 2016 | 8 museum experts and creative designers |

## The NSE experiences

Since its first version in 2016, the NSE has been applied in a variety of contexts and used for the creation of 27 experiences in 16 museums and sites, as for example (Roussou et al, 2017), (Mirashrafi, 2017), (Katifori et al, 2019a), (Katifori et al, 2019b) and (Economou et al, 2019).

Based on the Technology Readiness Level (TRL) scale, defined as a method of estimating technology maturity (Heder, 2017), we define a digital Experience Readiness Level (ERL), as shown in Table 2, to help us classify the digital experiences created with NSE.

Table 2. Experience Readiness Level scale and number of experiences created

| | **Experience readiness level** | **No of exp.** |
|---|---|---|
| 1 | Draft concept validated | |
| 2 | Experience structure validated | 2 |
| 3 | Experience structure and sample scenes validated in a simulated environment | 1 |
| 4 | Draft production (TTS may have been used for audio) validated in a simulated environment | 3 |
| 5 | Draft production validated on site by invited users | 15 |
| 6 | Final production demonstrated and evaluated on site (invited users) | 3 |
| 7 | Final production demonstrated and evaluated on site (public) | 7 |

An encouraging result of the practice of the tool in different contexts was that in all cases the produced experiences were appropriate for the particular cultural site and evaluation activities showed that the visitor feedback was indeed very positive.

The application and use of the Mobile Player in many different contexts helped us to very quickly identify most of the usability issues, that where resolved leasing to a stable and intuitive app.

The effectiveness of the Mobile Player helped us focus on the evaluation of the experiences themselves and thus identify more general and conceptual issues inherent in the design of digital storytelling experiences. These results are being analysed with the objective to produce a concrete set of guidelines for experience design to complement our authoring tool.

An interesting relevant find to be reported is the fact that the authors, given minimal guidelines as to the type of experience to design, pushed the boundaries of the tool with alternative uses of activity templates and unforeseen story structure approaches, resulting in a variety of experience types.

Table 3 provides an overview of the types of experiences created with the NSE, the majority of which have reached ERL 5 or higher, being full experience prototypes validated with invited visitors on site.

Table 3. Overview of NSE experiences grouped by type

| Experience type | Description | No. of exp. |
| --- | --- | --- |
| Multimedia guide | Multimedia experience offering information. Similar to a traditional audio guide. | 4 |
| Digital storytelling | Digital storytelling experience with a linear plot. Choices offered for accessing information content. | 12 |
| Interactive digital storytelling | Digital storytelling experience with a non-linear plot. Choices offered that affect the story line. | 3 |
| Dialogue-based social experience | Digital experience promoting dialogue-based interaction among users. | 1 |
| Experiential social experience | Digital experience that recreates concepts or procedures in a social context. | 2 |
| Gamified educational experience | Multimedia experience with storytelling and gamification elements. | 5 |

# Summary of findings

In this section we summarize the main findings of the on-going evaluation of the NSE and NMP. We focus mainly on issues related to the general authoring workflow and the conceptual difficulties the users encountered while creating storytelling experiences.

## Offering a smooth user experience

As a general conclusion, the Narralive Mobile Player in its latest version, as it has been gradually updated by the continuous evaluation process, is a very stable and efficient application for delivering digital storytelling content. Visitors are able to use the app with no instructions and with no reported usability issues. It offers an intuitive, smooth user experience and allows the users to focus on the content itself. Any criticism regarding the app's performance, had to do with the usability of specific experience templates used and this was taken into consideration for the next versions, to be improved both in the NMP and the NSE.

## An intuitive authoring tool

The general consensus of the authors for the NSE tool is that it is an "intuitive", "easy to use" tool both "for prototyping and also for producing the final experience" and additionally, "very stable". As observed so far, it is a tool with a steep learning curve, as in all cases the authors were able to use it, unsupervised, after a 1.5 - 2 hours hands-on tutorial and offering a comfortable user experience, as they were able to complete their work quickly and with minimal effort.

In all cases our team was available to support the authors with possible issues they experienced and authors were encouraged to take notes of their observations about all aspects of the tool and then discuss them with the evaluation team. Through this process we were able early on to resolve usability issues and improve the offered functionality as per the user needs. For example, the current functionality of the tool includes a preview possibility for a specific Chapter or the whole Story, an activity template allowing interactive exploration of an image as well as templates to prescribe the use of GPS, NFC or QR code based choices to define the progress of the story; all this functionality requested in early Workshops and evaluation activities.

However, users suggested that the tool would benefit from a set of tutorials, possibly in video form, to be used as reference to help them complete specific authoring tasks and improve their understanding of the use of the available templates.

## A tool for story scripting, prototyping or production?

The Storyboard Editor was initially conceived as a story prototyping tool, to support authors without particular technical expertise to define their story concept and structure and put the draft scenario and script of the story in this structure. However, with almost no exception, in none of the use cases was this put to practice. All authors started to brainstorm mostly on paper, then conceived the initial story structure and script in a text editor like Microsoft Word, and once there was a more or less stable draft they then moved to the NSE to prototype and test their experience. This workflow, although initially comfortable to the users, led in some cases to inconsistencies between the initial textual script and the experience in the NSE as changes in the story had to be maintained in two different editors

The tool allowed them to work during the prototyping phase in an intuitive way and be able to easily follow an iterative design approach with several cycles of design, evaluation and re-design. Taking into account the specific targeted user profile, priority was given to ease of use and the templates that would simplify the production of the story without the user having to resort to programming to synthesize the complete experience. The initial priority was the simpler templates, producing story Pages combining image, text and sound, assets that require minimum production. The authors were able very easily to add draft sketches or images and sound, draft recordings or text to speech.

Gradually however, since the resulting experience itself was stable and sleek, it became evident that the users felt the need to intervene more on the produced result and fine-tune different aspects like the experience visual design and layout, including colors, fonts and other visual elements, thus influencing the development of NSE towards a tool that would support a fully produced, working interactive experience, not just a prototype for design purposes. In this sense, the NSE and Mobile are now available to produce final experiences.

## Designing for interaction with the Site

Designing mobile experiences for cultural heritage implies a deep understanding of the constraints of the site itself, that being either a museum space or archaeological site. These may prescribe specific natural routes in the site for the visitor, taking into account possible chronological, thematical or other itineraries as well as distance constraints when visiting big sites. The SBE, being mainly a tool introduced after the initial experience conceptualization phase, implies that the experts have already taken into account these factors for the story concept and design.

As already discussed, heritage experts and story designers used mostly text editors to draft the story concept, also coupled with image editors or tools like Microsoft PowerPoint to design on a map or sketch of the site and situate the various parts of the experience on specific locations. It is amongst our future plans for extending the Editor to add support for this stage of the experience design.

### Adding more interactivity in the templates

In the context of the same aforementioned user requirement for experience production with the NSE, the users expressed the need to be able to author and add to their experiences more interactive activities. Several ideas were offered, all of which included a more active role on the side of the user. Some examples mentioned were:

- Augmented reality activities that enhance the exhibits in space
- VR activities that show exhibits in their actual context of use or in their original form
- Possibility for the users to add their own photos
- Possibility for the users to comment
- Being able to author timed events that will take place during the experience at pre-designed points in time or after another specific event has taken place.

These templates are being considered for the next version of the tool.

### The need to simplify the story structure

A consistent finding amongst most of the authors was the fact that it was not immediately clear what is the role of Chapters and Scenes and how they impact the created mobile experience. Although Chapters and Scenes are there to support the structuring of the content, they seem to create an overhead for the author, who does not necessarily wish to structure the experience in this way. Furthermore, it was not evident for the authors how this structure is reflected in the mobile user experience, and if the visitor will be able to see in any way the Chapters and Scenes. Additionally, the possibility to add preview images for the Chapters and Scenes to offer a better overview of the Story for the author added to this confusion of how these elements are to be made available to the visitors in the digital experience.

We plan to address this issue in the second version of our tool, by simplifying the story structure. The chapters will be removed as structural elements and only scenes (or "parts") will be maintained as grouping elements for the pages. This simpler structure foreseeing story "parts" and "links" between them will be more concise and at the same time allow for the creation of a wider variety of story graphs.

### The need to support complex branching narratives

A storyboard look-and-feel allows the author to take in the whole narrative flow with a single glance and can directly represent the linear paths that the user will experience. This approach was very efficient for the creation of mainly linear story structures.

However, branching seems also to be an important part of digital storytelling and the concept of branching and its visualization was probably the most challenging aspect of the tool. As branching complexity increases, it becomes more and more difficult for the author to visualize and to manage the structure. This is the reason some commented that they would have liked the possibility to have different overview options for the whole story structure in different levels of detail, as needed, to be able to view the story branches and zoom into specific nodes.

# Conclusions

The NSE has been used extensively in many different contexts by authors of varying experience and expertise and in various team combinations and sizes (cultural heritage experts, storytellers, museology students, museum educators, digital artists, e.tc.), and the end results ranged from single-day workshop productions, to student projects in museology classes, to professional multi-month undertakings, including a fully-produced storytelling experience offered as the official app of the University of Athens History Museum in Plaka, Athens, Greece.

The use of the tool so far has proven its advantages for effective design and development of a variety of digital mobile experiences. The storyboard paradigm by itself is not well-suited to very complex non-linear narrative structures, but may successfully represent near-linear structures (i.e., a main storyline with parallel paths (branches) that merge back to the main one). This type of narrative is well suited in a realistic setting where a cultural heritage site is looking to produce a cost-efficient storytelling experience, and an expert familiar with non-linear narratives is not part of the team.

Our observation that authors never seemed to start working directly on the tool but rather used a text editor as their tool of choice for concept design and scripting of the experience, led us to the design and development of a new tool to support this first step of the creative process. The tool is currently under development and soon its testing will commence. This design tool will combine a text editor look and feel with a graph-based visualization of the story structure, to address the reported structure overview issue.

Similarly, the evaluation identified a need for a supporting tool for the experts to plan how the experience unfolds in space, by connecting specific experience parts to specific locations. The design, however, of the experience itself can only be guided by the NSE and any other supporting tool. No tool can substitute the knowledge and expertise of the site experts on its spatial constraints. To this end, the offered tools should be coupled with detailed guidelines on how to approach these constraints even from the first steps of the experience design. We have worked towards a first version of these guidelines that can be found in the EMOTIVE project report D5.1 Conceptual Framework and Guide.

Apart from recording the authors' feedback, we also evaluated most experiences with visitors in the context of the NMP assessment. The results of this process are being analysed to help us better understand the function of the various activities and story elements and their combination in the context of mobile experiences for heritage. This analysis hints to important issues, some of the already identified in the literature as challenges not resolved, like finding the right balance between fiction and facts, story and the surrounding environment, entertainment and learning goals and it is our objective through the use of the NSE in many use cases to shed light to these issues and produce useful guidelines for engaging digital storytelling in heritage sites.

# Acknowledgements

This work was partly supported by the project EMOTIVE. EMOTIVE has received funding from the Horizon 2020 EU Framework Programme for Research and Innovation under grant agreement n° 727188.